\documentclass[hidelinks,12pt, letter paper]{article}
\usepackage[utf8]{inputenc}
\usepackage{authblk}
\usepackage{caption}
\usepackage{float}
\usepackage{subcaption}
\usepackage{graphicx}
\usepackage[hidelinks]{hyperref}
\usepackage{amsmath}
\usepackage{setspace}
\usepackage{array}
\usepackage{tikz}
\usepackage{tikz-feynman}
\usepackage{breqn}
\usepackage{cite}

\title {Stochastic parametric modulation of linear and non-linear oscillators: Perturbation theory of the response function}
\author{Sourin Dey \thanks{E-mail: sourin.dey71@gmail.com} $^1$ }
\author{Jayanta K. Bhattacharjee\thanks{E-mail: jayanta.bhattacharjee@gmail.com} $^2$ }
\affil{$^1$School of Physical Sciences, Indian Association for the Cultivation of Science, Jadavpur, Kolkata, 700032, India}
\affil{$^2$Department of Physics, Indian Institute of Technology, Kanpur, 208016, India}
\date{}

\numberwithin{equation}{section}

\begin{document}

\maketitle
\vspace{6pt}

\begin{abstract}
		We study a stochastically driven, damped nonlinear oscillator whose frequency is modulated by a white or coloured noise. Using diagrammatic perturbation theory, we find that in the absence of nonlinearity, parametric modulation by a coloured noise can lead to a Kapitza pendulum-like stabilization of an unstable configuration provided the noise is anti-correlated. Further, we show that for modulation by a white noise of amplitude $\lambda$ and correlation strength $F$, the system will have an extremely large response if the product of $\lambda^{2}F$ equals a specific combination of the frequency and the damping coefficient. This prediction can be experimentally tested.
\end{abstract}

\newpage

\section{Introduction}\label{Sec:Intro}

Stochastically driven oscillators have been the subject of extensive investigations over the last three decades. The effect of stochasticity on different kinds of attractors in dynamical systems has been studied in detail along with uses like the decay of metastable states, oscillations between multiple states, etc \cite{sagues2007spatiotemporal, perez2023universal,  maclaurin2022stochastic, wang1999stochastic, jung1993periodically, markus1988stochastic, dybiec2017underdamped, saha2007girsanov, saha2007higher, lifshitz2008nonlinear, sarkar2022method}. Recent work involving diagrammatic perturbation theory of statistical physics\cite{pal2023perturbation} has allowed a systematic inclusion of a variety of non-linear effects. Previously, nonlinear terms used to be handled by a self-consistent single-loop approximation. In this work, we extend the diagrammatic perturbation theory to the stochastic oscillator whose natural frequency is modulated by a stochastic forcing, i.e. there is a stochastic parametric modulation in the system.\\

The system is governed by the dynamics represented by,
\begin{align}\label{system_equ}
    \ddot{x}+2\Gamma \dot{x} +\omega_{0}^{2}(1+\lambda f(t)) x+\mu x^{3}=g(t)
\end{align}
In the above equation, we have restricted the nonlinearity to the cubic term. The inclusion of other nonlinearities will follow similar lines to what we will describe here. The driving stochastic forcing $g(t)$ will be taken to be a Gaussian white noise with the two-time correlation prescribed by,
\begin{align}\label{Noise-g-corr}
    \langle g(t_{1})g(t_{2}) \rangle =2D\delta(t_{1}-t_{2})
\end{align}
where $D$ is a constant.\\

The modulating term $f(t)$ on the other hand can be oscillatory(a variation on the Kapitza problem) or stochastic. Our primary concern is with a random $f(t)$. Once again the distribution of the stochastic $f(t)$ will be taken to be Gaussian. The noise $f(t)$ can, however, be white or coloured. The two-time correlation will be,
\begin{align}\label{Noise-f-corr}
    \langle f(t_{1})f(t_{2}) \rangle= 2F \delta(t_{1}-t_{2})
\end{align}
for the white noise modulation and
\begin{align}\label{Noise-f-corr-col}
     \langle f(t_{1})f(t_{2}) \rangle= \frac{2F}{\tau}e^{-|t_{1}-t_{2}|/\tau}
\end{align}
for the coloured noise. The prefactor F in the equation above is a constant. The limit $\tau\rightarrow 0$ in Eq. (\ref{Noise-f-corr-col}) corresponds to the white noise. We note that an oscillatory $f(t)=Fcos(\Omega t)$ with a very high $\Omega$ is a Kapitza-like situation. The diagrammatic expansion in powers of $\mu$($\lambda=0$) was dealt with in \cite{pal2023perturbation}. Here we show that if $\mu=0$, the effect of $\lambda$ can be cast as a similar diagrammatic perturbation theory and for $\lambda\neq0$, $\mu\neq0$, we generate a double power series with mixed terms of the form $\mu^{a}\lambda^{b}$ where $a$ and $b$ are integers. \\

In Sec. [\ref{section-two}], we will deal exclusively with the situation $\lambda\neq0$, $\mu=0$, and in Sec. [\ref{section-three}] we treat the case where $\lambda$ and $\mu$ are both non-zero. An interesting outcome of the linear but stochastically modulated system ($\mu=0$) is that at O($\lambda^{2}$) there is no contribution to the frequency shift if $f(t)$ is a white noise. On the other hand, if $f(t)$ is a coloured noise, then it is possible to have a shift in the natural frequency of the system in a manner very similar to the Kapitza pendulum. Furthermore, it will be seen that the standard nonlinear correction to the response can be strongly affected by the modulation terms.\\

Before finishing the introductory section, we would like to introduce the basic features of the calculation. Our focus will primarily be on the response function $R(\omega)$ defined by, 

\begin{align}\label{Response_defn}
    R(\omega)\delta(\omega+\omega')=\left\langle \frac{\delta x(\omega)}{\delta g(\omega')}\right\rangle
\end{align}

and the correlation function defined as

\begin{align}
    C(\omega)\delta(\omega+\omega')=\langle x(\omega)x(\omega')\rangle
\end{align}

The angular brackets in $R(\omega)$ and $C(\omega)$ imply that the correlation function is averaged over the noise terms $f(\omega)$ and $g(\omega)$. For calculational purposes, it is more convenient to write Eq. (\ref{Response_defn}) as 

\begin{align}\label{response_practical}
    R(\omega)\delta(\omega+\omega')=\frac{1}{2D}\langle x(\omega)g(\omega')\rangle
\end{align}

At any order of perturbation theory $R(\omega)$ will require two subscripts i.e.

\begin{align}
	\begin{split}
    R(\omega)&=\sum_{m,n}R_{mn}\mu^{m}\lambda^{n}\\
	&=R_{00}(\omega)+\mu R_{10}(\omega)+\lambda R_{01}(\omega)+\mu\lambda R_{11}(\omega)+\mu\lambda^{2}R_{12}(\omega)+...
	\end{split}
\end{align}

Since it is convenient to work in the frequency space we will rewrite Eq. (\ref{system_equ}) as

\begin{align}\label{system-eq-ft}
\begin{split}
    \left(\omega_{0}^{2}-\omega^{2}+2i\omega\Gamma\right)x(\omega)=-\lambda\omega_{0}^{2}\int x(\omega_{1})f(\omega-\omega_{1})\frac{d\omega_{1}}{2\pi}\\-\mu\int x(\omega_{1})x(\omega_{2})x(\omega-\omega_{1}-\omega_{2})\frac{d\omega_{1}}{2\pi}\frac{d\omega_{2}}{2\pi}+g(\omega)
\end{split}
\end{align}

The diagrammatic perturbation theory for $\lambda=0$ was extensively studied in \cite{pal2023perturbation}. Here, we need to expand the variable $x(\omega)$ in a double power series

\begin{align}\label{eq:position-expansion}
	\begin{split}
    x(\omega)&=\sum_{m,n}x_{mn}\mu^{m}\lambda^{n}\\
	&=x_{00}(\omega)+\mu x_{10}(\omega)+\lambda x_{01}(\omega)+\mu\lambda x_{11}(\omega)+\mu\lambda^{2}x_{12}(\omega)+...
	\end{split}
\end{align}

At the zeroth order ($\lambda=\mu=0$) we can read off the response function from Eq. (\ref{system-eq-ft}) and (\ref{response_practical})
as 
\begin{align}
    R_{00}(\omega)=\frac{1}{(\omega_{0}^{2}-\omega^{2}+2i\omega\Gamma)}
\end{align}
The corresponding correlation function is 
\begin{align}
    C_{00}(\omega)=\frac{2D}{(\omega_{0}^{2}-\omega^{2})^{2}+4\omega^{2}\Gamma^{2}}
\end{align}

For the nonlinear oscillator ($\mu\neq0$) it is well known that the response function can be significantly affected. The $O(\mu)$ correction to the frequency $\omega_{0}^{2}$ is given by $3\mu D/2\Gamma\omega_{0}^{2}$ from \cite{pal2023perturbation}. We will show that the $\mu\neq0$ effect is actually strongly modified by a parametric modulation. \\

We end this introductory section by considering a linear periodically modulated system where dynamics is given by
\begin{align}
    \ddot{x}+2\Gamma\dot{x} +\omega_{0}^{2}(1+\lambda\cos{\Omega t})x=g(t)
\end{align}
In frequency space, 
\begin{align}\label{kapitza_freq_space}
    \left(\omega_{0}^{2}-\omega^{2}+2i\omega\Gamma\right)x(\omega)=g(\omega) -\frac{\lambda\omega_{0}^{2}}{2}[x(\omega-\Omega)+x(\omega+\Omega)]
\end{align}
In a perturbative approach, we expand
\begin{align}
    x=x_{0}(\omega)+\lambda x_{1}(\omega)+\lambda^{2} x_{2}(\omega)+ ...
\end{align}
Inserting this in Eq. (\ref{kapitza_freq_space}), we obtain at different orders:

\begin{subequations}
    \begin{equation}\label{kapitza_order-sep-1}
        O(1): R_{0}^{-1}(\omega)x_{0}(\omega)=g(\omega) 
    \end{equation}
    \begin{equation}\label{kapitza_order-sep-2}
        O(\lambda): R_{0}^{-1}(\omega)x_{1}(\omega)=-\frac{\omega_{0}^{2}}{2}[x_{0}(\omega-\Omega)+x_{0}(\omega+\Omega)]
    \end{equation}
    \begin{equation}\label{kapitza_order-sep-3}
        O(\lambda^{2}): R_{0}^{-1}(\omega)x_{2}(\omega)=-\frac{\omega_{0}^{2}}{2}[x_{1}(\omega-\Omega)+x_{1}(\omega+\Omega)]
    \end{equation}
\end{subequations}

where $R_{0}^{-1}=(\omega_{0}^{2}-\omega^{2}+2i\omega\Gamma$) is the zeroth order response function. Adding all the above equations, Eq. (\ref{kapitza_order-sep-1}) - (\ref{kapitza_order-sep-3}), we find 

\begin{align}
    \begin{split}
        R_{0}^{-1}(\omega)[x_{0}(\omega)&+\lambda x_{1}(\omega)+\lambda^{2}x_{2}(\omega)]=g(\omega)-\frac{\omega_{0}^{2}\lambda}{2}[x_{0}(\omega-\Omega)+x_{0}(\omega+\Omega)]\\&+\frac{\lambda^{2}\omega_{0}^{4}}{4}[x_{0}(\omega)(R_{0}(\omega-\Omega)+R_{0}(\omega+\Omega))\\&+R_{0}(\omega-\Omega)x_{0}(\omega-2\Omega)+R_{0}(\omega+\Omega)x_{0}(\omega+2\Omega)]+...
    \end{split}
\end{align}

We can rewrite the above equation, correct to $O(\lambda^{2})$ in the very high frequency limit($\Omega>>\omega$) as 

\begin{align}
    \begin{split}
        R_{0}^{-1}(\omega)x(\omega)-\frac{\lambda^{2}\omega_{0}^{4}}{4}[R_{0}(\Omega)+R_{0}(-\Omega)]x(\omega)=&g(\omega)-\frac{\lambda\omega_{0}^{2}}{2}[x_{0}(\Omega)+x_{0}(-\Omega)]\\+\frac{\lambda^{2}\omega_{0}^{4}}{4}[x_{0}(2\Omega)R_{0}(\Omega)+x_{0}(-2\Omega)R_{0}(\Omega)]\\
     =&\bar{g}(\omega)
     \end{split}
\end{align}

Since, $R_{0}(\Omega)\approx -1/\Omega^{2}$ for very high $\Omega$. we can write the above dynamics, correct to $O(\lambda^{2})$, as 

\begin{align}
    \left[\omega_{0}^{2}+\frac{\lambda^{2}\omega_{0}^{4}}{2\Omega^{2}}-\omega^{2}+2i\Gamma\omega\right]x(\omega)=[\bar{\omega}_{0}^{2}-\omega^{2}+2i\Gamma\omega]=\bar{g}(\omega)
\end{align}

where $\bar{\omega_{0}^{2}}$ is an effective frequency

\begin{align}\label{Kapitza-freq-corr}
\bar{\omega_{0}^{2}}=\omega_{0}^{2}+\frac{\lambda^{2}\omega_{0}^{4}}{2\Omega^{2}}
\end{align}

and $\bar{g}(\omega)$ is an effective drive. If $\omega_{0}^{2}$
is negative as would be the case with an unstable situation, Eq. (\ref{Kapitza-freq-corr}) indicates that $\bar{\omega}_{0}^{2}$ can become positive for appropriate values of $\lambda$, $\omega_{0}$ and $\Omega$. This is equivalent to the well-known Kapitza effect.



\section{Linear oscillator with parametric modulation}\label{section-two}

We now switch on the parametric noise term and keep $\mu=0$. This means that we are looking at a linear but stochastically modulated system. This implies that we can write Eq. (\ref{system-eq-ft}) to

\begin{align}
\begin{split}
    \left(\omega_{0}^{2}-\omega^{2}+2i\omega\Gamma\right)x(\omega)=-\lambda\omega_{0}^{2}\int x(\omega_{1})f(\omega-\omega_{1})\frac{d\omega_{1}}{2\pi}+g(\omega)
\end{split}
\end{align}

We write the expansion of Eq. (\ref{eq:position-expansion}) in the form,

\begin{align}
	x(\omega)=x_{00}(\omega)+\lambda x_{01}(\omega)+ \lambda^{2} x_{02}(\omega)+ ...
\end{align}
At different orders of $\lambda$ we obtain,

\begin{align}
	O(\lambda^{0}): \;  (\omega_{0}^{2}-\omega_{2}+2i\Gamma\omega)x_{00}(\omega)&=g(\omega)\label{eq:order-zero}
\end{align}

This defines the zeroth order response function
\begin{align}\label{eq:zeroth-order-response}
	R_{00}^{-1}(\omega)=(\omega_{0}^{2}-\omega_{2}+2i\Gamma\omega)
\end{align}
and 
\begin{align}
	O(\lambda): \; (\omega_{0}^{2}-\omega_{2}+2i\Gamma\omega)x_{01}(\omega)&=-\omega_{0}^{2}\int f(\omega_{1})x_{00}(\omega-\omega_{1})\frac{d\omega_{1}}{2\pi}\label{eq:order-lambda} \\
	O(\lambda^{2}):\; (\omega_{0}^{2}-\omega_{2}+2i\Gamma\omega)x_{02}(\omega)&=-\omega_{0}^{2}\int f(\omega_{1})x_{01}(\omega-\omega_{1})\frac{d\omega_{1}}{2\pi}\label{eq:order-lambda-sq}
\end{align}

Our aim is to construct the response function as defined in Eq. (\ref{response_practical}). At the zeroth order, the response function is given in Eq. (\ref{eq:zeroth-order-response}). At the next order, the response function will be given by 
\begin{align}
	R_{01}(\omega)&=\frac{1}{2D}\frac{\langle x_{01}(\omega)g(\omega')\rangle}{\delta(\omega+\omega')}
\end{align}

Using Eq. (\ref{eq:order-lambda}), the correlation in the numerator has the expression, 

\begin{align}
	\langle x_{01}(\omega)g(\omega')\rangle&=-\left< \omega_{0}^{2}R_{00}(\omega)\int x_{00}(\omega_{1})f(\omega-\omega_{1})\frac{d\omega_{1}}{2\pi}g(\omega')\right>
\end{align}

Since, from Eq. (\ref{eq:order-zero}), we know that $x_{00}(\omega)$ is proportional to $g(\omega)$ which correlation with $f(\omega-\omega_{1})$, the quantity will vanish implying
\begin{align}
	R_{01}(\omega)=0
\end{align}

We move on to the next order, $O(\lambda^{2})$ and construct $\langle x_{02}(\omega)g(\omega') \rangle$ from Eq. (\ref{eq:order-lambda-sq}) to obtain

\begin{align}\label{eq:lambda-sq-step}
\begin{split}
    \langle x_{02}(\omega)g(\omega')\rangle
    =\omega_{0}^{4}(2D)R_{00}(\omega)\int R_{00}(\omega-\omega_{1}-\omega_{2})R_{00}(\omega-\omega_{1})\\\langle f(\omega_{1})f(\omega_{2})\rangle\delta(\omega'+\omega-\omega_{1}-\omega_{2})\frac{d\omega_{1}}{2\pi}\frac{d\omega_{2}}{2\pi}
\end{split}
\end{align}
To proceed further we need to prescribe the nature of noise $f$. If we assume it to be white noise, then Eq. (\ref{Noise-f-corr}) holds. With this prescription, the response function $R_{02}(\omega)$ according to Eq. (\ref{response_practical}) from the above equation will be
 
\begin{align}
\begin{split}
    R_{02}(\omega)=&\omega_{0}^{4}(2F)R_{00}^{2}(\omega)\int R_{00}(\omega-\omega_{1}) \frac{d\omega_{1}}{2\pi}\\
    =&0
\end{split}
\end{align}
We thus find that for a white noise modulation of the linear restoring force, there is no contribution to the response function at the second order in perturbation theory. Given that the non-linear term that we are dealing with really has a very simple structure, this is not very unexpected. We now investigate whether a coloured noise is capable of producing a finite correction at this virtually trivial order of perturbation theory.\\

Accordingly, we consider the noise $f$ to be described by Eq. (\ref{Noise-f-corr-col}). In Fourier space, we have the same relation expressed as
\begin{align}
	\langle f(\omega)f(\omega')\rangle=&\frac{4F}{1+\omega^{2}\tau^{2}}\delta(\omega+\omega')-\frac{2F\tau}{(1-i\omega\tau)(1+i\omega'\tau)}
\end{align}

The second-order correction is now obtained by using the same, instead of the white noise form in Eq. (\ref{eq:lambda-sq-step}). This yields

\begin{align}
\begin{split}\label{eq:colored-response-order-lambda-squared}
         \langle x_{02}(\omega)g(\omega')\rangle
         &=\omega_{0}^{4}(2D)(2F) R_{00}(\omega) \int R_{00}(-\omega')R_{00}(\omega-\omega_{1}) \left[ \frac{2}{1+\omega_{1}^{2}\tau}\delta(\omega+\omega')\right. \\ 
 & \left. {} -\tau^{2}\frac{1}{(1-i\omega_{1}\tau)(1+i(\omega+\omega'-\omega_{1})\tau)} \right]\frac{d\omega_{1}}{2\pi}
\end{split}
\end{align}

For convenience of calculation, we will assume that the noise is weakly coloured which means that we take a small-$\tau$ limit. As is evident in Eq. (\ref{Noise-f-corr-col}), the limit $\tau\rightarrow0$ corresponds to white noise. For small but finite $\tau$, the response function $R_{02}(\omega)$, from the above equation, will be given by

\begin{align}
	\begin{split}
	R_{02}(\omega)&=\omega_{0}^{4}(2F)R_{00}^{2}(\omega)\int R_{00}(\omega-\omega_{1})\frac{2}{1+\omega_{1}^{2}\tau^{2}}\frac{d\omega_{1}}{2\pi}\\
	&=4F\omega_{0}^{4}R_{00}^{2}(\omega)\int \frac{1}{(\omega_{0}^{2}-(\omega-\omega_{1})^{2}+2i\Gamma(\omega-\omega_{1}))}\frac{1}{(1+\omega_{1}^{2}\tau^{2})}\frac{d\omega_{1}}{2\pi}
	\end{split}
\end{align}

If $R(\omega)$ is written as $R^{-1}(\omega)=\omega_{0}^{2}+\Delta\omega_{0}^{2}-\omega^{2}+2i\Gamma\omega$ then expanding in $\Delta\omega_{0}^{2}$, we see that $R(\omega)=(\omega_{0}^{2}-\omega^{2}+2i\omega\Gamma)-\Delta\omega_{0}^{2}/(\omega_{0}^{2}-\omega^{2}+2i\omega\Gamma)^{2}$. This identifies
\begin{align}
		\Delta\omega_{0}^{2}=-\frac{2\pi\omega_{0}^{4}F}{1+2\Gamma\tau+\omega_{0}^{2}\tau^{2}}
\end{align}
as the shift in the natural frequency.

We end this section by recalling the diagrammatic representation of the perturbation theory, which will enable us to write down the changes in the response function at different orders efficiently. We use the following notation for the basic elements entering the perturbation theory.
\begin{table}[h]
\centering
\begin{tabular}{wl{2cm} wl{1cm} |wl{6cm}}
	\begin{tikzpicture}
		\begin{feynman}
			\vertex (a) at (0,0);
			\vertex (b) at (2,0);
			\diagram* {
				(a) -- [plain, thick] (b),
			};	
		\end{feynman}
	\end{tikzpicture}
	& &
	$x_{0}(\omega)$\\
	\begin{tikzpicture}
		\begin{feynman}
			\vertex (a) at (0,0);
			\vertex (b) at (2,0);
			\diagram* {
				(a) -- [boson, thick] (b),
			};
		\end{feynman}
	\end{tikzpicture}
	 & &
	$f(\omega)$	\\
	\begin{tikzpicture}
		\begin{feynman}
			\vertex (a) at (0,0.5);
			\vertex (b) at (1,1);
			\vertex (c) at (1,0);
			\vertex (d) at (0.5,0.5);
			\diagram* {
				(d) -- [boson, thick] (c),
				(d) -- [plain, thick] (b),
				(a) -- [plain, thick] (d)
			};
			\filldraw[fill=white, draw=black] (d) circle (3pt);
		\end{feynman}
	\end{tikzpicture}
	& &
	Vertex $\lambda$\\
	\begin{tikzpicture}
		\begin{feynman}
			\vertex (a) at (0,0);
			\vertex (b) at (1,0);
			\vertex (c) at (0,1);
			\vertex (d) at (1,1);
			\vertex (e) at (0.5, 0.5);
			\diagram* {
				(a) -- [plain, thick] (e),
				(b) -- [plain, thick] (e),
				(c) -- [plain, thick] (e),
				(d) -- [plain, thick] (e),
			};
			\fill (e) circle (3pt);
		\end{feynman}
	\end{tikzpicture}
	 & &
	Vertex $\mu$ \\
	\begin{tikzpicture}
		\begin{feynman}
			\vertex (a) at (0,0);
			\vertex (b) at (1,0);
			\vertex (c) at (2,0);
			\diagram* {
				(a) -- [plain, thick] (b),
				(b) -- [boson, thick] (c),
			};
		\end{feynman}
	\end{tikzpicture}
 			& & 
	Response function, $R(\omega)$\\
\end{tabular}
	\caption{Elements of diagrammatic perturbation theory}
\end{table}
\\

Using the above elements, the contribution of Eq. (\ref{eq:lambda-sq-step}) to the response function can be written as
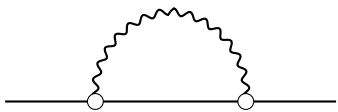
\begin{figure}[H]
	\centering
	\begin{tikzpicture}
	\begin{feynman}
		\vertex (a) at (-0.2,0);
		\vertex (b) at (1,0);
		\vertex (c) at (3,0);
		\vertex (d) at (4.2,0);
		\vertex (e) at (2,1.2);

		\diagram* {
			(a) -- [plain, thick] (b),
			(b) -- [plain, thick] (c),
			(c) -- [plain, thick] (d),
			(b) -- [boson, quarter left, thick] (e),
			(c) -- [boson, quarter right, thick] (e)
		};
		\filldraw[fill=white, draw=black] (b) circle (3pt);
		\filldraw[fill=white, draw=black] (c) circle (3pt);
	\end{feynman}
\end{tikzpicture}
	\caption{Lowest non-trivial response function}
	\label{fig:order-lambda-squared}
\end{figure}
The contribution of Fig.(\ref{fig:order-lambda-squared}) to the response function is exactly what is obtained in Eq. (\ref{eq:lambda-sq-step}). It vanishes for a white noise and is non-zero for a coloured noise and leads to Eq. (\ref{eq:colored-response-order-lambda-squared}). The very significant outcome of this section lies in the fact that a parametric modulation of the linear system by a coloured noise actually alters the response function of the system. The qualitative effect of a coloured noise is consequently similar to that of a non-linear term(quadratic or quartic) which can alter the response of the system. The unexpected feature is that a correlated noise can destabilize the system while an anti-correlated noise can stabilize it. The traditional ways of handling parametrically modulated systems can be found in \cite{rugar1991mechanical, kawai2002parametrically,butikov2004parametric,kourdis2006some,mathew2016dynamical,han2016stochastic,sheth2019noise,belousov2020volterra}.



\section{Non-linear oscillator with parametric modulation}\label{section-three}

In this section, we investigate whether the parametric modulation can affect the contribution to the frequency shift brought about by the nonlinear term. To do that we note that the expansion in Eq. (\ref{eq:position-expansion}) is the following.
\begin{align}
	\begin{split}
	x(\omega)&=x_{00}(\omega)+\mu x_{10}(\omega)+\mu^{2}x_{20}(\omega)+...\\
			& +\lambda[x_{01}(\omega)+\mu x_{11}(\omega)+\mu^{2}x_{21}(\omega)+...]\\ 
& +\lambda^{2}[x_{02}\omega)+\mu x_{12}(\omega)+\mu^{2}x_{22}(\omega)+...]+...
	\end{split}
\end{align}
 We then use the above expansion in Eq. (\ref{system-eq-ft}) and extract the relevant terms of mixed orders. Here we would be looking at $O(\mu\lambda)$ and $O(\mu \lambda^{2})$ terms. At $O(\mu\lambda)$ we have,

\begin{align}
	\begin{split}
    x_{11}(\omega)&=-\omega_{0}^{2}R_{00}(\omega)\int f(\omega')x_{10}(\omega-\omega')\frac{d\omega'}{2\pi}\\
&-3R_{00}(\omega)\int x_{01}(\omega_{1})x_{00}(\omega_{2})x_{00}(\omega-\omega_{1}-\omega_{2})\frac{d\omega_{1} d\omega_{2}}{(2\pi)^{2}}
	\end{split}
\end{align}

As per routine we look at the quantity $\langle x_{11}(\omega)g(\omega')\rangle$. It will have two terms corresponding to the two on the right-hand side of the above equation. The first term contains $x_{10}(\omega)$ which is the $O(\mu)$ term. We note that $x_{10}$ will not contain any $f$s because only orders with non-zero $\lambda$s can bring in $f$s. We can therefore conclude that the first term vanishes due to averaging over a single $f(\omega)$. The second term meets the same fate but due to averaging over an odd number of $g(\omega)$ s. Therefore, the response function at this order, $R_{11}(\omega)$, does not exist. Different approaches to handle weak nonlinearities can be found in \cite{vincent2021vibrational} and \cite{ducimetiere2022weakly}.\\

Moving on to $O(\mu\lambda^{2})$, extracting the expression for $x(\omega)$ at this order yields

\begin{align}\label{equation-order-lambda-squared-mu}
\begin{split}
    x_{12}(\omega)&=-\omega_{0}^{2}R_{00}(\omega)\int f(\omega')x_{11}(\omega-\omega')\frac{d\omega'}{2\pi}\\
    &-3R_{00}(\omega)\int x_{02}(\omega_{1})x_{00}(\omega_{2})x_{00}(\omega-\omega_{1}-\omega_{2})\frac{d\omega_{1}}{2\pi}\frac{d\omega_{2}}{2\pi}\\
    &-3R_{00}(\omega)\int x_{01}(\omega_{1})x_{01}(\omega_{2})x_{00}(\omega-\omega_{1}-\omega_{2})\frac{d\omega_{1}}{2\pi}\frac{d\omega_{2}}{2\pi}
\end{split}
\end{align}

\begin{figure}[h]
	\begin{subfigure}[t]{0.31\textwidth}
		\begin{tikzpicture}
			\begin{feynman}
			\vertex (a) at (0,0);
			\vertex (b) at (0.8,1.5);
			\vertex (e) at (2,0);
			\vertex (d) at (2,1.5);
			\vertex (c) at (2,3);
			\vertex (f) at (3.2,1.5);
			\vertex (g) at (4,0);
			\diagram* {
				(a) -- [plain, thick] (e),
				(e) -- [plain, thick] (g),
				(e) -- [plain, quarter left, looseness=1.0, thick] (b),
				(b) -- [plain, quarter left, looseness=1.0, thick] (c),
				(c) -- [plain, quarter left, looseness=1.0, thick] (f),
				(f) -- [plain, quarter left, looseness=1.0, thick] (e),
				(b) -- [boson, thick] (d),
				(d) -- [boson, thick] (f)
			};
		\end{feynman}
			\fill (e) circle (3pt);
			\filldraw[fill=white, draw=black] (b) circle (3pt);
			\filldraw[fill=white, draw=black] (f) circle (3pt);
	\end{tikzpicture}
	\caption{}
	\label{fig:order-mu-lambda-squared-1}
	\end{subfigure}%
	\begin{subfigure}[t]{0.32\textwidth}
	\begin{tikzpicture}
		\begin{feynman}
			\vertex (a) at (0.5,0);
			\vertex (b) at (0.9,2);
			\vertex (p) at (0.9,1);
			\vertex (e) at (2,0);
			\vertex (d) at (3.9,0);
			\vertex (q) at (4.2,0);
			\vertex (c) at (2,3);
			\vertex (f) at (3.1,2);
			\vertex (l) at (3.6, 1);
			\vertex (r) at (3.1,1);
			\vertex (g) at (5,0);
			
			\diagram *{
				(a) -- [plain, thick] (e),
				(e) -- [plain, thick] (g),
				(e) -- [plain, quarter left, looseness=0.7, thick] (p),
				(b) -- [plain, quarter left, looseness=0.7, thick] (c),
				(c) -- [plain, quarter left, looseness=0.7, thick] (f),
				(r) -- [plain, quarter left, looseness=0.7, thick] (e),
				(p) -- [plain, quarter left, looseness=0.3, thick] (b),
				(r) -- [plain, quarter right, looseness=0.3, thick] (f),
				(f) -- [boson, thick] (l),
				(l) -- [boson, thick] (q),
			};
		\end{feynman}
			\fill (e) circle (3pt);
			\filldraw[fill=white, draw=black] (f) circle (3pt);
			\filldraw[fill=white, draw=black] (q) circle (3pt);
	\end{tikzpicture}
		\caption{}
		\label{fig:order-mu-lambda-squared-2}
	\end{subfigure}%
	\begin{subfigure}[t]{0.31\textwidth}
		\centering
	\begin{tikzpicture}
		\begin{feynman}
			\vertex (a) at (0.5,0);
			\vertex (b) at (2,0);
			\vertex (c) at (3.5,0);
			\vertex (d) at (2,3);
			\vertex (e) at (1,0.8);
			\vertex (f) at (1,2.2);
			\vertex (g) at (2,1.5);
			\diagram *{
				(a) -- [plain, thick] (b),
				(b) -- [plain, thick] (c),
				(b) -- [plain, quarter left, looseness=0.7, thick] (e),
				(e) -- [plain, quarter left, looseness=0.7, thick] (f),
				(f) -- [plain, quarter left, looseness=0.7, thick] (d),
				(b) -- [plain, half right, thick] (d),
				(e) -- [boson, quarter right, looseness=0.7, thick] (g),
				(g) -- [boson, quarter right, looseness=0.7, thick] (f)
				};
			\fill (b) circle (3pt);
			\filldraw[fill=white, draw=black] (e) circle (3pt);
			\filldraw[fill=white, draw=black] (f) circle (3pt);
		\end{feynman}
	\end{tikzpicture}
	\caption{}
	\label{fig:order-mu-lambda-squared-3}
	\end{subfigure}%
	\caption{}
	\label{fig:order-mu-lambda-squared}
\end{figure}

The first two terms on the right-hand side of Eq. \ref{equation-order-lambda-squared-mu} give rise to disconnected diagrams except one, Fig. (\ref{fig:order-mu-lambda-squared-3}), arising from the second term. The third term is of primary interest as it corresponds to the non-reducible diagrams shown in Figs. (\ref{fig:order-mu-lambda-squared-1}) and (\ref{fig:order-mu-lambda-squared-2}). 
\\

We'll first find out the contribution of the first diagram to the quantity $\langle x_{12}(\omega)g(\bar{\omega})\rangle$. For that, we recall the form of $x_{01}(\omega)$ and insert it in the third term on the RHS of Eq. (\ref{equation-order-lambda-squared-mu}) to obtain

\begin{align}
    \begin{split}
        \langle x_{12}(\omega)g(\bar{\omega})\rangle=
        -3\omega_{0}^{4}R_{00}(\omega)\int R_{00}(\omega_{1})R_{00}(\omega_{2})\langle f(\omega')f(\omega'')\rangle \\
        \langle x_{00}(\omega_{1}-\omega')  x_{00}(\omega_{2}-\omega'')\rangle  \langle x_{00}(\omega-\omega_{1}-\omega_{2})g(\bar{\omega})\rangle \frac{d\omega_{1}}{2\pi}\frac{d\omega_{2}}{2\pi}\frac{d\omega'}{2\pi}\frac{d\omega''}{2\pi}
    \end{split}
\end{align}

Considering $f(t)$ to be white, the response function at this order, $R_{12}^{a}(\omega)$ coming from Fig. (\ref{fig:order-mu-lambda-squared-2}) will be

\begin{align}
    \begin{split}
         R_{12}^{a}(\omega)&=
        -3\omega_{0}^{4}2F R_{00}^{2}(\omega)\int R_{00}(\omega_{1})R_{00}(-\omega_{1})\frac{d\omega_{1}}{2\pi}\int C_{00}(\omega_{1}-\omega')\frac{d\omega'}{2\pi}\\
        &=-6\omega_{0}^{4}F R_{00}^{2}(\omega)\left(\frac{1}{4\Gamma \omega_{0}^{2}}\right)\left(\frac{2D}{4\Gamma\omega_{0}^{2}}\right)
    \end{split}
\end{align}
The correction to the natural frequency at this point is therefore
\begin{align}
	\bar{\omega}_{0}^{2}=\omega_{0}^{2}+\mu\frac{3D}{2\Gamma\omega_{0}^{2}}+\mu\lambda^{2}\frac{3FD}{4\Gamma^{2}}
\end{align}
We now find the contribution from the second diagram Fig.  (\ref{fig:order-mu-lambda-squared-2}). 

\begin{align}
    \begin{split}
         \langle x_{12}(\omega)g(\bar{\omega})\rangle=
         -3\omega_{0}^{4}R_{00}(\omega)\int R_{00}(\omega_{1})R_{00}(\omega-\omega_{1}-\omega_{2})
         \langle f(\omega')f(\omega'')\rangle \\ \langle x_{00}(\omega_{2})x_{00}(\omega_{1}-\omega'')\rangle \langle x_{00}(\omega-\omega_{1}-\omega_{2}-\omega_{3})g(\bar{\omega})\rangle \frac{d\omega_{1}}{2\pi}\frac{d\omega_{2}}{2\pi}\frac{d\omega'}{2\pi}\frac{d\omega''}{2\pi}
    \end{split}
\end{align}

As before, considering $f$ to be white and contracting the delta functions, the response function takes the form

\begin{align}\label{eq:second-diag-contribution}
    \begin{split}
        R_{12}^{b}(\omega)&=-6\omega_{0}^{4}F R_{00}^{2}(\omega)\int R_{00}(\omega_{1})R_{00}(\omega-\omega_{1}-\omega_{2})\frac{d\omega_{1}}{2\pi}\int C_{00}(\omega_{2})\frac{d\omega_{2}}{2\pi}\\
	&=-6\omega_{0}^{4}F R_{00}^{2}(\omega)\int J(\omega, \omega_{2})C_{00}(\omega_{2})\frac{d\omega_{2}}{2\pi}\\
	&=-6\omega_{0}^{4}F R_{00}^{2}(\omega)I(\omega)\\
    \end{split}
\end{align}

where,
\begin{align}\label{eq:I}
	I(\omega)=\int J(\omega, \omega_{2})C_{00}(\omega_{2})\frac{d\omega_{2}}{2\pi}
\end{align}
and
\begin{align}\label{eq:J}
		J(\omega, \omega_{2})&=J(\omega-\omega_{2})=\int R_{00}(\omega_{1})R_{00}(\omega-\omega_{1}-\omega_{2}) \frac{d\omega}{2\pi}\\
	&=\int R_{00}(\omega_{1})R_{00}(\Omega-\omega_{1})\frac{d\omega}{2\pi}
\end{align}
with $\Omega=\omega-\omega_{2}$. This makes it clear that the correction $R_{12}^{b}(\omega)$ will depend on the frequency ($\omega$) in a non-trivial fashion and hence the response function $R(\omega)$ will no longer have the form $R^{-1}(\omega)=\bar{\omega}_{0}^{2}+2i\Gamma\omega-\omega^{2}$ with $\bar{\omega}_{0}^{2}$ a constant. Instead, we know
\begin{align}
	\bar{\omega}_{0}^{2}=\omega_{0}^{2}+\frac{3}{2}\frac{\mu D}{\Gamma\omega_{0}^{2}}+\frac{3}{4}\mu\lambda^{2}\frac{FD}{\Gamma^{2}}+6\omega_{0}^{4}F I_{R}(\omega)
\end{align}
with $I_{R}(\omega)$ being the real part of $I(\omega)$ given in Eq. (\ref{eq:I}).
\\

We end this section by pointing out that we can very easily write down all contributions of $O(\mu\lambda^{2n})$ where $n$ is an integer(contributions of $O(\mu\lambda^{2n+1})$ vanish). This implies that if $\mu$ is a rather small number and $\lambda$ significantly greater than $\mu$ but not necessarily unity, then all terms of $O(\mu\lambda^{2n})$ contribute a summable series. The corresponding answers are given in Sec. \ref{section-four}.



\section{Contribution from a class of higher order diagrams}\label{section-four}

In the previous section, we looked at the two loop diagrams of $O(\mu\lambda^{2})$ which are shown in Fig. (\ref{fig:order-mu-lambda-squared-1})-(\ref{fig:order-mu-lambda-squared-3}). There are actually an infinite number of closely related diagrams shown in Fig. (\ref{fig:series-one}) and (\ref{fig:series-two}).\\

\begin{figure}[h]

	\begin{subfigure}[t]{0.31\textwidth}
            \centering
	\begin{tikzpicture}[xscale=0.8, yscale=0.8]
		\begin{feynman}
			\vertex (a) at (0.5,0);
			\vertex (b) at (0.8,2);
			\vertex (p) at (0.8,1);
			\vertex (e) at (2,0);
			\vertex (d) at (2,2);
			\vertex (q) at (2,1);
			\vertex (c) at (2,3);
			\vertex (f) at (3.2,2);
			\vertex (r) at (3.2,1);
			\vertex (g) at (3.5,0);
			
			\diagram *{
				(a) -- [plain, thick] (e),
				(e) -- [plain, thick] (g),
				(e) -- [plain, quarter left, looseness=1.0, thick] (p),
				(b) -- [plain, quarter left, looseness=1.0, thick] (c),
				(c) -- [plain, quarter left, looseness=1.0, thick] (f),
				(r) -- [plain, quarter left, looseness=1.0, thick] (e),
				(p) -- [plain, quarter left, looseness=0.3, thick] (b),
				(r) -- [plain, quarter right, looseness=0.3, thick] (f),
				(b) -- [boson, thick] (d),
				(d) -- [boson, thick] (f),
				(p) -- [boson, thick] (q),
				(q) -- [boson, thick] (r),
			};
		\end{feynman}
			\fill (e) circle (3pt);
			\filldraw[fill=white, draw=black] (b) circle (3pt);
			\filldraw[fill=white, draw=black] (f) circle (3pt);
			\filldraw[fill=white, draw=black] (p) circle (3pt);
			\filldraw[fill=white, draw=black] (r) circle (3pt);
	\end{tikzpicture}
		\caption{}
		\label{fig:order-mu-lambda-squared-a}
	\end{subfigure}%
+
	\begin{subfigure}[t]{0.31\textwidth}
            \centering
	\begin{tikzpicture}[xscale=0.8, yscale=0.8]
		\begin{feynman}
			\vertex (a) at (0.5,0);
			\vertex (b) at (0.8,2.4);
			\vertex (p) at (0.8,1.0);
			\vertex (s) at (0.7, 1.7);
			\vertex (e) at (2,0);
			\vertex (d) at (2,2.4);
			\vertex (q) at (2,1.0);
			\vertex (t) at (2,1.7);
			\vertex (c) at (2,3.5);
			\vertex (f) at (3.2,2.4);
			\vertex (r) at (3.2,1.0);
			\vertex (u) at (3.3, 1.7);
			\vertex (g) at (3.5,0);
			
			\diagram *{
				(a) -- [plain, thick] (e),
				(e) -- [plain, thick] (g),
				(e) -- [plain, quarter left, looseness=1.0, thick] (p),
				(b) -- [plain, quarter left, looseness=1.0, thick] (c),
				(c) -- [plain, quarter left, looseness=1.0, thick] (f),
				(r) -- [plain, quarter left, looseness=1.0, thick] (e),
				(p) -- [plain, quarter left, looseness=0.3, thick] (s),
				(r) -- [plain, quarter right, looseness=0.3, thick] (u),
				(s) -- [plain, quarter left, looseness=0.3, thick] (b),
				(u) -- [plain, quarter right, looseness=0.3, thick] (f),
				(b) -- [boson, thick] (d),
				(d) -- [boson, thick] (f),
				(p) -- [boson, thick] (q),
				(q) -- [boson, thick] (r),
				(s) -- [boson, thick] (t),
				(t) -- [boson, thick] (u),
			};
		\end{feynman}
			\fill (e) circle (3pt);
			\filldraw[fill=white, draw=black] (b) circle (3pt);
			\filldraw[fill=white, draw=black] (f) circle (3pt);
			\filldraw[fill=white, draw=black] (p) circle (3pt);
			\filldraw[fill=white, draw=black] (r) circle (3pt);
			\filldraw[fill=white, draw=black] (s) circle (3pt);
			\filldraw[fill=white, draw=black] (u) circle (3pt);
	\end{tikzpicture}
	\caption{}
	\label{fig:order-mu-lambda-squared-b}
	\end{subfigure}%
+ ...
		\caption{}
	\label{fig:series-one}
\end{figure}

The contributions from the Fig. (\ref{fig:order-mu-lambda-squared-a}) and (\ref{fig:order-mu-lambda-squared-b}) are found to be $3\mu\lambda^{4}F^{2}D\omega_{0}^{2}/8\Gamma^{3}$ and $3\mu\lambda^{6}F^{3}D\omega_{0}^{3}/16\Gamma^{4}$ respectively. Consequently, the dressed frequency $\bar{\omega}_{0}$ has the relation,
\begin{align}\label{eq:series-one-sum}
	\bar{\omega}^{2}=\omega_{0}^{2}+\frac{3}{2}\frac{\mu D}{\Gamma\omega_{0}^{2}}+\frac{3}{4}\mu\lambda^{2}\frac{FD}{\Gamma^{2}}+\frac{3}{8}\mu\lambda^{4}\frac{F^{2}D\omega_{0}^{2}}{\Gamma^{3}}+\frac{3}{16}\mu\lambda^{6}\frac{F^{3}D\omega_{0}^{4}}{\Gamma^{4}}+...
\end{align}

Recognizing the emerging G.P. series in Eq. (\ref{eq:series-one-sum}), the expression for $\bar{\omega}_{0}$ after summing the series is given by

\begin{align}
	\bar{\omega}_{0}^{2}=\omega_{0}^{2}+\frac{3}{2}\frac{\mu D}{\Gamma\omega_{0}^{2}}+\frac{3}{4}\frac{\frac{\mu \lambda^{2}DF}{4\Gamma}}{1-\frac{\lambda^{2}F\omega_{0}^{2}}{2\Gamma}}
\end{align}

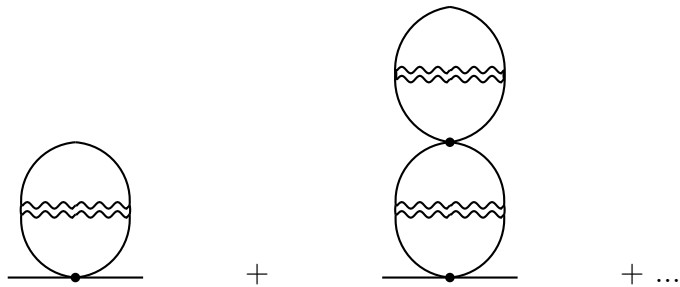
\begin{figure}[h]
    \begin{subfigure}[t]{0.33\textwidth}	
	\centering	
    \begin{tikzpicture}[xscale=0.6, yscale=0.6]
		\begin{feynman}
			\vertex (a) at (0.5,0);
			\vertex (b) at (0.8,1.6);
			\vertex (p) at (0.8,1.4);
			\vertex (e) at (2,0);
			\vertex (d) at (2,1.6);
			\vertex (q) at (2,1.4);
			\vertex (c) at (2,3);
			\vertex (f) at (3.2,1.6);
			\vertex (r) at (3.2,1.4);
			\vertex (g) at (3.5,0);
			
			\diagram *{
				(a) -- [plain, thick] (e),
				(e) -- [plain, thick] (g),
				(e) -- [plain, quarter left, looseness=1.0, thick] (p),
				(b) -- [plain, quarter left, looseness=1.0, thick] (c),
				(c) -- [plain, quarter left, looseness=1.0, thick] (f),
				(r) -- [plain, quarter left, looseness=1.0, thick] (e),
				(p) -- [plain, quarter left, looseness=0.3, thick] (b),
				(r) -- [plain, quarter right, looseness=0.3, thick] (f),
				(b) -- [boson, thick] (d),
				(d) -- [boson, thick] (f),
				(p) -- [boson, thick] (q),
				(q) -- [boson, thick] (r),
			};
		\end{feynman}
			\fill (e) circle (3pt);
	
\end{tikzpicture}
	\end{subfigure}%
+
	\begin{subfigure}[t]{0.33\textwidth}
	\centering
		\begin{tikzpicture}[xscale=0.6, yscale=0.6]
		\begin{feynman}
			\vertex (a) at (0.5,0);
			\vertex (b) at (0.8,1.6);
			\vertex (p) at (0.8,1.4);
			\vertex (e) at (2,0);
			\vertex (d) at (2,1.6);
			\vertex (q) at (2,1.4);
			\vertex (c) at (2,3);
			\vertex (f) at (3.2,1.6);
			\vertex (r) at (3.2,1.4);
			\vertex (g) at (3.5,0);
			\vertex (l) at (2,6);
			\vertex (j) at (0.8, 4.4);
			\vertex (h) at (0.8, 4.6);
			\vertex (k) at (3.2, 4.4);
			\vertex (i) at (3.2, 4.6);	
			\vertex (m) at (2,4.4);		
			\vertex (n) at (2,4.6);
			\diagram *{
				(a) -- [plain, thick] (e),
				(e) -- [plain, thick] (g),
				(e) -- [plain, quarter left, looseness=1.0, thick] (p),
				(b) -- [plain, quarter left, looseness=1.0, thick] (c),
				(c) -- [plain, quarter left, looseness=1.0, thick] (f),
				(r) -- [plain, quarter left, looseness=1.0, thick] (e),
				(p) -- [plain, quarter left, looseness=0.3, thick] (b),
				(r) -- [plain, quarter right, looseness=0.3, thick] (f),
				(c) -- [plain, quarter left, looseness=1.0, thick] (h),
				(j) -- [plain, quarter left, looseness=1.0, thick] (l),
				(l) -- [plain, quarter left, looseness=1.0, thick] (k),
				(i) -- [plain, quarter left, looseness=1.0, thick] (c),
				(h) -- [plain, quarter left, looseness=0.3, thick] (j),
				(i) -- [plain, quarter right, looseness=0.3, thick] (k),
				(b) -- [boson, thick] (d),
				(d) -- [boson, thick] (f),
				(p) -- [boson, thick] (q),
				(q) -- [boson, thick] (r),
				(j) -- [boson, thick] (m),
				(m) -- [boson, thick] (k),
				(h) -- [boson, thick] (n),
				(n) -- [boson, thick] (i),
			};
		\end{feynman}
			\fill (e) circle (3pt);
			\fill (c) circle (3pt);
\end{tikzpicture}
	\end{subfigure}%
+ ...
    \caption{Series with higher orders of $\mu$ where the first diagram depicts the sum of the series given in Fig. (\ref{fig:series-one}). }
    \label{fig:high-mu-order}
\end{figure}

The singularity at $\lambda^{2}F\omega_{0}^{2}=2\Gamma$ is an unexpected feature. However, it is not of any concern because when one carries out the sum over all possible diagrams present in the series shown in Fig. (\ref{fig:high-mu-order}), it is ensured that the validity of the perturbation series is extended to $\lambda^{2}F\omega_{0}^{2}/2\Gamma\sim 1/\mu$ and hence the response is large but not infinite. Experimental investigations should reveal a large shift of the natural frequency when this condition is satisfied. This modulation-induced large shift of the natural frequency at small values of the non-linearity parameter is the principal result of our work.\\

\begin{figure}[h]
	\begin{subfigure}[t]{0.4\textwidth}
            \centering
		\begin{tikzpicture}[xscale=0.8, yscale=0.8]
		\begin{feynman}
			\vertex (a) at (0.5,0);
			\vertex (b) at (0.9,2);
			\vertex (p) at (0.9,1);
			\vertex (e) at (2,0);
			\vertex (d) at (3.9,0);
			\vertex (q) at (4.6,0);
			\vertex (c) at (2,3);
			\vertex (f) at (3.1,2);
			\vertex (r) at (3.1,1);
			\vertex (g) at (5.5,0);
			
			\diagram *{
				(a) -- [plain, thick] (e),
				(e) -- [plain, thick] (g),
				(e) -- [plain, quarter left, looseness=0.7, thick] (p),
				(b) -- [plain, quarter left, looseness=0.7, thick] (c),
				(c) -- [plain, quarter left, looseness=0.7, thick] (f),
				(r) -- [plain, quarter left, looseness=0.7, thick] (e),
				(p) -- [plain, quarter left, looseness=0.3, thick] (b),
				(r) -- [plain, quarter right, looseness=0.3, thick] (f),
				(f) -- [boson, thick] (q),
				(r) -- [boson, thick] (d),
			};
		\end{feynman}
			\fill (e) circle (3pt);
			\filldraw[fill=white, draw=black] (f) circle (3pt);
			\filldraw[fill=white, draw=black] (r) circle (3pt);
			\filldraw[fill=white, draw=black] (d) circle (3pt);
			\filldraw[fill=white, draw=black] (q) circle (3pt);
	\end{tikzpicture}
	\caption{}
	\label{fig:series-2-a}
	\end{subfigure}%
+
	\begin{subfigure}[t]{0.4\textwidth}
            \centering
			\begin{tikzpicture}[xscale=0.8, yscale=0.8]
		\begin{feynman}
			\vertex (a) at (0,0);
			\vertex (b) at (0.9,2.4);
			\vertex (p) at (0.9,1.0);
			\vertex (s) at (0.8, 1.7);
			\vertex (e) at (2,0);
			\vertex (d) at (3.9,0);
			\vertex (t) at (4.5,0);
			\vertex (q) at (5.1,0);
			\vertex (c) at (2,3.5);
			\vertex (f) at (3.1,2.6);
			\vertex (r) at (3.1,1.0);
			\vertex (u) at (3.2, 1.7);
			\vertex (g) at (6,0);
			
			\diagram *{
				(a) -- [plain, thick] (e),
				(e) -- [plain, thick] (g),
				(e) -- [plain, quarter left, looseness=0.7, thick] (p),
				(b) -- [plain, quarter left, looseness=0.7, thick] (c),
				(c) -- [plain, quarter left, looseness=0.7, thick] (f),
				(r) -- [plain, quarter left, looseness=0.7, thick] (e),
				(p) -- [plain, quarter left, looseness=0.3, thick] (s),
				(r) -- [plain, quarter right, looseness=0.3, thick] (u),
				(s) -- [plain, quarter left, looseness=0.3, thick] (b),
				(u) -- [plain, quarter right, looseness=0.3, thick] (f),
				(r) -- [boson, thick] (d),
				(u) -- [boson, thick] (t),
				(f) -- [boson, thick] (q),
			};
		\end{feynman}
			\fill (e) circle (3pt);
			\filldraw[fill=white, draw=black] (d) circle (3pt);
			\filldraw[fill=white, draw=black] (f) circle (3pt);
			\filldraw[fill=white, draw=black] (q) circle (3pt);
			\filldraw[fill=white, draw=black] (r) circle (3pt);
			\filldraw[fill=white, draw=black] (t) circle (3pt);
			\filldraw[fill=white, draw=black] (u) circle (3pt);
			\filldraw[fill=white, draw=black] (d) circle (3pt);
			\filldraw[fill=white, draw=black] (q) circle (3pt);
			\filldraw[fill=white, draw=black] (t) circle (3pt);
			
	\end{tikzpicture}
	\caption{}
	\label{fig:series-2-b}
	\end{subfigure}%
 + ...
		\caption{}
	\label{fig:series-two}
\end{figure}

We now focus on the second set of diagrams shown in Fig. (\ref{fig:series-two}). The contribution from the diagram in Fig. (\ref{fig:series-2-a}) is seen to be 

\begin{align}
	\begin{split}
	R_{14}(\omega)=-3\omega_{0}^{8}(2F)^{2}R_{00}^{2}(\omega)\int R(\omega_{1})&R(\omega_{3})R(\omega-\omega_{1}-\omega_{2})R(\omega-\omega_{1}+\omega_{3})\\
	 &C_{00}(\omega_{2}) \frac{d\omega_{1}}{2\pi}	\frac{d\omega_{2}}{2\pi}\frac{d\omega_{3}}{2\pi}
	\end{split}
\end{align}

However, we find the following 
\begin{align}\label{eq:vanishing}
	\int R(\omega_{3})R(\omega-\omega_{1}+\omega_{3})\frac{d\omega_{3}}{2\pi}=0
\end{align}
hence, the contribution from Fig. (\ref{fig:series-2-a}) vanishes. Moreover, the contribution from the subsequent diagrams in the series in Fig. (\ref{fig:series-two}) contains the vanishing integral in Eq.(\ref{eq:vanishing}) in increasing powers. Therefore, we find no contribution from the diagrams of this series.

\section{Conclusion}
We have dealt with a stochastically driven oscillator subjected to a stochastic parametric modulation of the linear restoring force. It was found that for a coloured noise modulation, 
the linear system can have a  Kapitza-like stabilization of an unstable fixed point if the noise term modulating the linear restoring force is coloured and anti-correlated. We have also shown that if the modulating random force has a white noise correlation, then the coupling between the modulation and non-linearity can have striking effects which can make the oscillator show very large responses at critical values of the system parameters-the natural frequency, the damping and the strength of the noise-correlations.

\bibliographystyle{ieeetr}
\bibliography{references.bib}

\end{document}